\documentclass[prc,twocolumn,epsfig,nofootinbib,floatfix,showpacs]{revtex4}
\usepackage{graphics}
\usepackage{epsfig}
\usepackage{amsfonts}
\usepackage{amsmath}
\usepackage{bm}

\usepackage{color}

\begin{document}
\title{Low-energy $M1$ states in deformed nuclei: spin-scissors or spin-flip?}
\author{V. O. Nesterenko$^{1,2,3}$, P. I. Vishnevskiy$^{1,4}$, A. Repko$^{5}$, J. Kvasil$^{6}$}
\affiliation{$^1$ Laboratory of Theoretical
Physics, Joint Institute for Nuclear Research, Dubna, Moscow region, 141980, Russia}
\affiliation{$^2$ State University "Dubna", Dubna, Moscow Region 141980, Russia}
\affiliation{$^3$ Moscow Institute of Physics and Technology, Dolgoprudny, Moscow Region 141701, Russia}
\email{nester@theor.jinr.ru}
\affiliation{$^4$ Institute of Nuclear Physic Almaty, Almaty Region, Kazakhstan}
\affiliation{$^5$ Institute of Physics, Slovak Academy of Sciences, 84511, Bratislava, Slovakia}
\affiliation{$^6$ Institute of Particle and Nuclear Physics, Charles
University, CZ-18000, Praha 8, Czech Republic}

\begin{abstract}
The low-energy $M1$ states in deformed $^{164}$Dy and spherical $^{58}$Ni
are explored in the framework of fully self-consistent Quasiparticle Random-Phase
Approximation (QRPA) with various Skyrme forces. The main attention is paid to
orbital and spin $M1$ excitations. The obtained results are compared with the
prediction of the low-energy {\it spin-scissors} $M1$ resonance suggested  within
Wigner Function Moments (WFM) approach. A possible relation of this resonance
to low-energy spin-flip excitations is analyzed. In connection with recent WFM studies,
we consider evolution of the low-energy spin-flip states in $^{164}$Dy with deformation
(from the equilibrium value to the spherical limit). The effect
of tensor forces is briefly discussed.  It is shown that two groups
of $1^+$ states observed at 2.4-4 MeV in $^{164}$Dy  are rather explained by
fragmentation of the orbital $M1$ strength than by the occurrence of the collective
spin-scissors resonance. In general, our calculations do not confirm the existence
of this resonance.
\end{abstract}


\maketitle

\section{Introduction}
\label{intro}

In addition to the familiar $M1$ orbital-scissors resonance (OSR)
\cite{Iud78,Boh84,Iud_PN97}, two specific
low-energy spin-scissors resonances (SSR) were predicted within the
macroscopic WFM approach, see e.g. \cite{Bal_NPA11,Bal_PRC15,Bal_PRC18,Bal_PRC,Bal_PAN}
and references therein.
While OSR is macroscopically treated as out-of-phase oscillations of proton
and neutron deformed subsystems (see Fig. 1-a), two SSRs are viewed as
out-of-phase oscillations of the deformed subsystems with different directions of
nucleon spins. As seen from Fig. 1-b,c, the first spin-scissors resonance
(SSR-1) represents oscillations of nucleons with spin-up against the nucleons with spin-down.
In the second spin-scissors resonance (SSR-2), the neutron and proton spins in each scissors
blade have opposite directions. So, following WFM approach, the nuclear scissors mode
should have a triple structure: OSR + two SSR branches. All the scissors states
should demonstrate significant $M1(K=1)$ transitions to the ground state.

 \begin{figure}[t] 
\centering
\includegraphics[width=6cm]{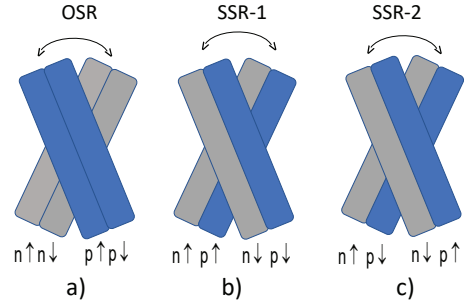}
\caption{Scissors triple \cite{Bal_PRC}:
OSR (a), SSR-1 (b), SSR-2 (c). The neutron (proton) axially deformed fractions
are shown by light (dark) bars. The spin direction of nucleons is indicated by arrows.
Each mode in the triple exhibits scissors-like oscillations of two blades:
neutrons vs protons in OSR, spin-up vs spin-down nucleons in SSR-1
(spins of neutrons and protons in each blade have the same direction), and SSR-2
oscillations where the neutron and proton spins in each blade have opposite directions.}
\label{fig:1}
\end{figure}

In our recent study \cite{Ne_PRC21}, we analyzed the WFM prediction of SSR in the framework
of  fully
self-consistent Quasiparticle Random-Phase Approximation (QRPA) method with
Skyrme forces \cite{Rep_arxiv,Rep_EPJA17,Rep_sePRC19,Kva_seEPJA19}. Our microscopic
calculations have shown that particular deformed nuclei, e.g. $^{160,162,164}$Dy, can indeed
exhibit at 2-3 MeV some  $K^{\pi}=1^+$ states with a noticeable $M1$ strength. These states
are formed by ordinary spin-flip transitions between spin-orbit partners
with a low orbital momentum. The states are not collective. In
principle, they could be associated with the predicted SSR. However, QRPA distributions
of the nuclear current in these states do not show any signatures of the spin-scissors flow.
Moreover, our calculations for $^{160,162,164}$Dy and $^{232}$Th reveal a strong interference
between orbital and spin contributions to $M1$ strength.
The low-energy $M1$ states turn out to be a mixture of the orbital and spin-flip modes.
As a result, the experimental data
in  $^{160,162,164}$Dy and $^{232}$Th \cite{Mar_exp_NRF_1995,Val_exp_2015,Ren_exp_Oslo_2018,Ade_232Th},
treated by WFM as a confirmation of SSR, can be mainly explained
by fragmentation of the orbital strength.

One of the main points disputed in QRPA \cite{Ne_PRC21} and WFM \cite{Bal_PAN} studies is the
scissors picture of the low-energy spin states. The WFM authors insist in the scissors scheme
\cite{Bal_PRC,Bal_PAN} and WFM distributions of the nuclear current indeed remind SSR-1 and SSR-2
modes shown in Fig. 1b,c. At the same time, in the last publications \cite{Bal_PRC,Bal_PAN} WFM authors
accept a possible spin-flip origin of SSR. By our opinion,
if we admit the spin-flip character of the low-energy of $K^{\pi}=1^+$,  then we should reject
the scissors scheme. Indeed, the scissors excitations by definition can exist only in deformed
nuclei while low-energy spin-flip states can occur in both deformed and spherical nuclei. The scissors
scheme obviously contradicts the existence of low-energy spin-flip states in spherical nuclei.

To inspect this point in more detail, the evolution of low-energy $M1$ strength with
decrease of the deformation parameter $\delta$ was recently studied within WFM for $^{164}$Dy
\cite{Bal_PAN}. The obtained results look strange: the $M1$ strength, both spin and orbital,
fully disappears at $\delta < 0.13 \div 0.15$.  This result looks natural for the orbital$M1$ strength
(which indeed has to vanish in the spherical limit) but unphysical for the spin-flip
$M1$ strength (which generally has no any reason to vanish in the spherical limit).

In this connection, we suggest here investigation of low-energy $M1$ strength in  $^{164}$Dy
in the framework of our fully self-consistent matrix QRPA method \cite{Rep_arxiv,Rep_EPJA17,Rep_sePRC19,Kva_seEPJA19}.  Various Skyrme forces
(SkM*~\cite{SkM*}, SG2~\cite{SG2}, and two SV-forces SVbas and SV-tls~\cite{SVbas}) are used.
First of all, we inspect evolution of low-energy
$M1$ strength in  $^{164}$Dy with decreasing nuclear quadrupole deformation.  We get
an expected result for the spherical limit: the orbital $M1$ strength
almost vanishes while the amount of spin $M1$ strength remains almost the same.
Besides, we check QRPA current distributions, including those averaged ones,
but do not get the current similar to SSR-1 and SSR-2. To demonstrate the
existence of low-energy spin-flip states in spherical nuclei, we calculate $M1$ strength
in $^{58}$Ni.

It is known that spin-flip states can be sensitive to tensor forces
\cite{Colo_tf,Ves_PRC09,Shen_PLB18}. So
we explore $M1$ strength in $^{164}$Dy with the Skyrme
parametrization SV-tls involving the tensor impact.

\section{Calculation scheme}

The calculations are performed within the self-consistent QRPA model
\cite{Rep_arxiv,Rep_EPJA17,Rep_sePRC19,Kva_seEPJA19} based on the Skyrme functional
\cite{Ben_RMP03}. The model is fully self-consistent since a) both mean field and residual
interaction are derived from the initial Skyrme functional b) the residual interaction
takes into account all the terms following from functional, c)  Coulomb (direct and
exchange) parts are involved, d) both particle-hole  and particle-particle  channels are included
\cite{Rep_EPJA17}.  The spurious admixtures caused by violation of the rotational invariance
are removed using the technique \cite{Kva_seEPJA19}.

The representative set of Skyrme forces is used. We employ
the standard force SkM* \cite{SkM*}, the force SG2 \cite{SG2} which is widely
used in the analysis of $M1$ excitations \cite{Ne_PRC21,Ves_PRC09,Nes_JPG10},
recently developed force SVbas \cite{SVbas} and its analog with the tensor contribution
SV-tls \cite{SVbas}. As seen from Table \ref{tab-1},
these forces have different isoscalar $b_4$ and isovector $b_4'$ spin-orbit parameters
(see their definitions in Refs. \cite{Ves_PRC09,Stone_PPNP07}). In SkM* and SG2,
the usual convection $b_4=b'_4$ is used.
In SVbas, the separate tuning of $b_4$ and $b_4'$ is done. The force  SV-tls is constructed like
SVbas but employs the full tensor spin-orbit terms. These Skyrme forces
reproduce the two-hump structure of $M1(K=1)$ spin-flip giant resonance in open-shell nuclei, see e.g.
\cite{Ne_PRC21,Ves_PRC09,Nes_JPG10}.

\begin{table} 
\centering
\caption{Isoscalar effective mass $m^*_0$,  isoscalar and isovector spin-orbit
parameters $b_4$ and $b'_4$, type of pairing, and parameter
$\beta$ of the equilibrium axial quadrupole deformation (vs the experimental
values  \cite{database}) for Skyrme forces SkM*, SG2, SVbas and SV-tls.}
\label{tab-1}       
\begin{tabular}{|c|c|c|c|c|c|c|}
\hline
 force       & $m^*_0$  & $b_4$ & $b'_4$ & pairing & $\beta$\\
 & &  MeV $\rm{fm}^5$      &  MeV $\rm{fm}^5$  &   &  \\
\hline
SkM* & \; 0.79 & 65.0 & 65.0 &  volume & 0.354\\
\hline
SG2  & \; 0.79 & 52.5 & 52.5 & volume & 0.354 \\
\hline
SVbas  & \; 0.90 & 62.32 & 34.11 & surface & 0.348 \\
\hline
SV-tls  & \; 0.90 & 62.32 & 0.0001 & surface & 0.344 \\
\hline
exper &&&&& 0.349(3)\\
\hline
\end{tabular}
\end{table}

For axially deformed $^{164}$Dy, the nuclear mean field and pairing are computed with the
code SKYAX  \cite{SKYAX} using a two-dimensional grid in cylindrical coordinates.
The calculation box extends up to three times the nuclear radius, the grid step is
1 fm. The axial quadrupole equilibrium deformation is obtained by minimization of
the energy of the system. As seen from Table \ref{tab-1}, the obtained values
of the deformation  parameters $\beta$  are in a good agreement with
the experimental data \cite{database}, especially for SVbas. The volume and surface
pairing is treated at the level of the iterative HF-BCS
(Hartree-Fock plus Bardeen-Cooper-Schrieffer) method \cite{Rep_EPJA17}.
To cope with divergent character of zero-range pairing forces, energy-dependent
cut-off factors are used, see for details Refs. \cite{Ne_PRC21,Rep_EPJA17,Be00}.
The calculations for spherical $^{58}$Ni are performed by a similar manner but using SKYAX version
for spherical nuclei.

Both QRPA codes for deformed and spherical nuclei are implemented in the matrix
form~\cite{Rep_arxiv}. A large configuration space is used.
The single-particle spectrum extends from the bottom of the potential well up to 30 MeV.
For example, in SkM* calculations for  $^{164}$Dy, we use 693 proton and 807 neutron
single-particle levels.  The two-quasiparticle (2qp) basis in QRPA calculation
for $K^{\pi}=1^+$ states includes 4882 proton and 9702 neutron configurations.

For axially deformed nuclei, the reduced probability for $M1(K=1)$ transitions ($M11$
in the short notation) from the ground state $|0\rangle$ to the excited QRPA state
$|\nu\rangle$ with $I^{\pi}K=1^+1$ reads
\begin{equation}
\label{eq:BM11}
B_{\nu}(M11)=2|\:\langle\nu|\:\hat{\Gamma}(M11)\:|0\rangle \:|^2 .
\end{equation}
The coefficient 2 means that contributions of both projections $K=$1 and -1
are taken into account.  The transition operator has the form
\begin{equation}
\label{eq:M1}
 {\hat \Gamma}(M11) =
 \mu_N \sqrt{\frac{3}{4\pi}}\sum_{q \epsilon p,n}
[g^{q}_s {\hat s}(\mu=1) + g^{q}_l {\hat l} (\mu=1)]
\end{equation}
where $\mu_N$ is the nuclear magneton, ${\hat{s}}(\mu=1)$ and ${\hat{l}}(\mu=1)$ are
$\mu$=1 projections of the standard spin and orbital operators,
$g^{q}_s$ and $g^{q}_l$
are spin and orbital gyromagnetic factors. We use the quenched spin g-factors
$g^{q}_s = \eta \bar{g}^{q}_s$ where  $\bar{g}^{p}_s$ = 5.58  and  $\bar{g}^{n}_s$ =-3.82
are bare proton and neutron g-factors and $\eta$=0.7 is the familiar
quenching parameter \cite{Har01}. The orbital g-factors are  $g^{p}_l$ = 1
and  $g^{n}_l$ = 0. In what follows, we consider three
relevant cases: {\it spin} ($g^{q}_l=0$), {\it orbital} ($g^{q}_s=0$),
and {\it total} (when both
spin and orbital transitions are taken into account).

In deformed $^{164}$Dy, the modes of multipolarities $M11$ and $E21$ are mixed.
So we also calculate the reduced probability
\begin{equation}
\label{BE21}
B_{\nu}(E21)=2|\:\langle\nu|\:\hat{\Gamma}(E21)\:|0\rangle \:|^2
\end{equation}
with the quadrupole transition operator
\begin{equation}
\label{eq:E21}
 {\hat \Gamma}(E21) =
  e \sum_{q \epsilon p,n} e_{\text{eff}}^q
r^2 Y_{21}(\theta,\phi)
\end{equation}
where $Y_{21}(\theta,\phi)$ is the spherical harmonic and $e_{\text{eff}}^q$ are
effective charges. Here we use $e_{\text{eff}}^p$=1 and $e_{\text{eff}}^n$=0.
For the sake of brevity, the notations $B(M11)$ and $B(E21)$ are below replaced
by the shorter notations $B(M1)$ and $B(E2)$.

We also calculate current transition densities (CTD)
\begin{equation}
\delta \bold {j}_{\nu}(\bold{r}) = \langle \nu| \hat{\bold j}|0\rangle (\bold{r})
\label{CTD}
\end{equation}
using operator of the convective nuclear current
\begin{equation}
\label{j_con}
\hat{\bold j} (\bold r)= -i \frac{e\hbar}{2m} \sum_{q =n,p}e_{\text{eff}}^q
\sum_{k \epsilon q}(\delta({\bold r} - {\bold r}_k) {\bold \nabla}_k
+ {\bold \nabla}_k \delta({\bold r} - {\bold r}_k)) .
\end{equation}
Here $e_{\text{eff}}^q$ are the effective charges. They are
$e_{\text{eff}}^p$=1 and $e_{\text{eff}}^n$=0 for the proton current,
$e_{\text{eff}}^p$=0 and $e_{\text{eff}}^n$=1 for the neutron current,
$e_{\text{eff}}^p=e_{\text{eff}}^n$=1 for isoscalar current and
$e_{\text{eff}}^p=-e_{\text{eff}}^n$=1 for isovector current.

In calculation of spin-up and spin-down CTD, we project
the QRPA wave function  $|\nu \rangle $
to the proper spin direction using spinor structure of the involved
single-particle wave functions in cylindric coordinates \cite{Nes_PRC06}.

For $^{58}$Ni, we use similar expressions relevant for spherical nuclei.

\begin{figure}
\includegraphics[width=8.5cm]{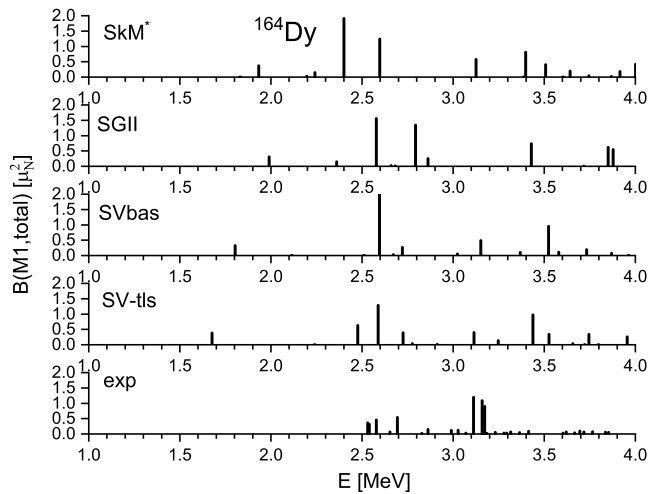}
\label{fig:2}
\caption{Reduced transition probabilities $B(M\text{1,total})$ in $^{164}$Dy, calculated with
the forces SkM*, SG2, SVbas, SV-tls. QRPA results  are compared with the experimental data
\cite{Mar_exp_NRF_1995}.}
\end{figure}

\begin{table}
\caption{Characteristics of particular low-energy $K^{\pi}=1^+$
states in $^{164}$Dy, calculated  within QRPA with the forces SkM*, SG2, SVbas and SV-tls.
For each state, we show the excitation energy E, orbital, spin
and total reduced transition probabilities $B(M1)$ and main 2qp components
(contribution to the state norm  in $\%$, structure in terms of Nilsson
asymptotic quantum numbers, position of the involved single-particle
states relative to the Fermi level F).}
\centering
\label{tab-2}
\begin{tabular}{|c|c|c|c|c|c|c|c|}
\hline\hline
force & E & \multicolumn{3}{c|}{$B(M1) [\mu^2_N]$} & \multicolumn{3}{c|}{Main 2qp components} \\
\hline
& [MeV] & orb & spin & total & $\%$  & $[N, n_z, \Lambda]$ & F-pos \\
\hline
SkM$^{*}$&  1.93 & 0.03 & 0.37 & 0.18 & 96 & $ pp[411\uparrow, 411\downarrow] $ & F-1, F+1 \\
&  & & & & 2 & $ nn[521\uparrow, 521\downarrow] $ & F-2, F+1 \\
&  2.24 & $\approx 0$ & 0.11 & 0.07 & 80 & $ nn[521\uparrow, 521\downarrow] $ & F-2, F+1 \\
&  & & & & 17 & $ nn[521\uparrow, 523\uparrow] $ & F-2, F \\
&  2.40 & 0.53 & 0.06 & 0.96 & 71 & $ nn[642\uparrow, 633\uparrow] $ & F-1, F+2 \\
&  & & & & 16 & $ pp[532\uparrow, 523\uparrow] $ & F-2, F+2 \\
&  2.60 & 0.25 & 0.08 & 0.62 & 56 & $ pp[532\uparrow, 523\uparrow] $ & F-2, F+2 \\
& & & & & 13 & $ nn[521\uparrow, 512\uparrow] $ & F-2,F+3 \\
\hline
SG2&  1.99 & 0.01 & 0.25 & 0.15 & 99 & $ pp[411\uparrow, 411\downarrow] $ & F, F+1 \\
&  2.36 & $\approx 0$ & 0.09 & 0.07 & 98 & $ nn[521\uparrow, 521\downarrow] $ & F-2, F+1 \\
&  2.58 & 0.42 & 0.05 & 0.78 & 73 & $ nn[642\uparrow, 633\uparrow] $ & F, F+2 \\
&  & & & & 11 & $ pp[532\uparrow, 523\uparrow] $ & F-1, F+2 \\
&  2.79 & 0.31 & 0.07 & 0.67 & 62 & $ pp[532\uparrow, 523\uparrow] $ & F-1, F+2 \\
&  & & & & 12 & $ nn[521\uparrow, 512\uparrow] $ & F-2, F+3 \\
\hline
SVbas&  1.80 & 0.03 & 0.32 & 0.16 & 99 & $ pp[411\uparrow, 411\downarrow] $ & F, F+1 \\
&  2.59 & 0.41 & 0.14 & 1.02 & 69 & $ pp[532\uparrow, 523\uparrow] $ & F-1, F+2 \\
&  & & & & 8 & $ nn[521\uparrow, 521\downarrow] $ & F-2, F+1 \\
&  2.72 & 0.10 & $\approx 0$ & 0.13 & 71 & $ nn[642\uparrow, 633\uparrow] $ & F-1, F+2 \\
&  & & & & 13 & $ nn[521\uparrow, 512\uparrow] $ & F-2, F+3 \\
\hline
SV-tls&  1.67 & 0.07 & 0.50 & 0.19 & 95 & $ pp[411\uparrow, 411\downarrow] $ & F, F+1 \\
&  & & & & 2 & $ nn[521\uparrow, 521\downarrow] $ & F-3, F+1 \\
&  2.48 & 0.17 & 0.02 & 0.31 & 52 & $ pp[532\uparrow, 523\uparrow] $ & F-1, F+2 \\
&  & & & & 25 & $ nn[521\uparrow, 521\downarrow] $ & F-3, F+1 \\
&  2.59 & 0.18 & 0.15 & 0.64 & 67 & $ nn[521\uparrow, 521\downarrow] $ & F-3, F+1 \\
&  & & & & 23 & $ pp[532\uparrow, 523\uparrow] $ & F-1, F+2 \\
\hline\hline
\end{tabular}
\\
\end{table}

\section{Results and discussion}
\subsection{$^{164}$Dy: $M1$ strength and structure of $1^+$ states}

In Fig. 2, the total (orbital+spin) $M1$ strength at 1-4 MeV in $^{164}$Dy, calculated with
the forces SkM*, SG2, SVbas and  SV-tls, is compared with the experimental data
\cite{Mar_exp_NRF_1995}. It is seen that all four Skyrme forces give,
in accordance with the experiment, a group of peaks around 2.5 MeV.
The second group
observed at 3.1-3.2 MeV can be roughly associated with the calculated peaks at
3.1-3.6 MeV. Below 2.4 MeV the calculations give a well separated peak located at
1.94 (SkM*), 1.98 (SG2), 1.81 (SVbas) and 1.68 (SV-tls) MeV. As shown below, this peak
has a spin-flip origin. Note that database \cite{database} suggests a tentative
$1^+$ state at 1.74 MeV. The comparison of SVbas and SV-tls results shows that impact
of tensor forces on low-energy $1^+$ states in $^{164}$Dy is rather weak.

In Table~\ref{tab-2}, we demonstrate features of some relevant $1^+$ states
in $^{164}$Dy, calculated with different Skyrme forces. For all the forces,
the lowest  $1^+$ state is dominated by two-quasiparticle (2qp) proton configuration
$pp[411\uparrow, 411\downarrow]$ and so is basically of a spin-flip character.
In this state, we obtain $B(M1,\text{spin}) > B(M1,\text{orbit})$. For SkM*
and SG2, the second $1^+$ state
is also spin-flip one but now with the dominant neutron component
 $ nn[521\uparrow, 521\downarrow] $. The origin of these proton and neutron 2qp
spin-flip components in Dy isotopes is discussed in detail in our previous study
\cite{Ne_PRC21}. Other states in Table~\ref{tab-2} are basically orbital
with $B(M1,\text{orbit}) > B(M1,\text{spin})$. These
states are more collective than spin-flip ones. The column "F-pos" shows that all
the considered excitations are formed by particle-hole ($1ph$) transitions.

It is easy to see that in most of the considered states
there is a significant interference of the spin and orbital contributions in the
total $M1$ transition. In other words,
$B(M1,{\text{spin}})+B(M1,\text{orbit}) \ne B(M1,\text{total})$.
The interference is usually destructive in spin-flip states and constructive in
orbital states. So, in all the states, the spin and orbital $M1$ modes are mixed.
WFM also predicts a mixture of OSR, SSR-1 and SSR-2 \cite{Bal_PRC,Bal_PAN}.

\subsection{$^{164}$Dy: current fields}

\begin{figure}
\includegraphics[width=8cm]{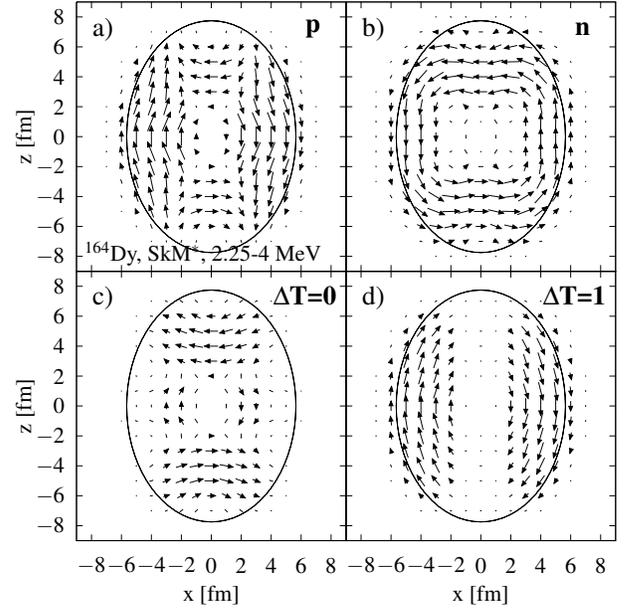}
\label{fig:3}
\caption{QRPA SkM* proton (a), neutron (b), isoscalar (c) and isovector (d)
distributions of the nuclear current (convective CTD on (x,z) plane), averaged
at the energy range 2.25-4 MeV in $^{164}$Dy. The solid ellipse shows the nuclear boundary.}
\end{figure}
\begin{figure}
\includegraphics[width=8cm]{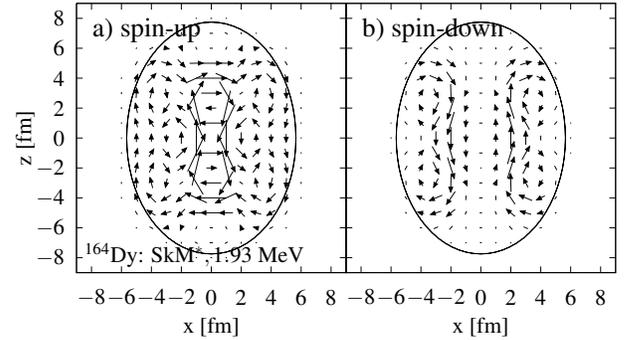}
\label{fig:4}
\caption{SkM* spin-up (a) and spin-down (b) CTD for the lowest QRPA $1^+$ state
at 1.93 MeV in $^{164}$Dy.  The solid ellipse shows the nuclear boundary.}
\end{figure}

\begin{figure*}[t]
\includegraphics[width=12cm]{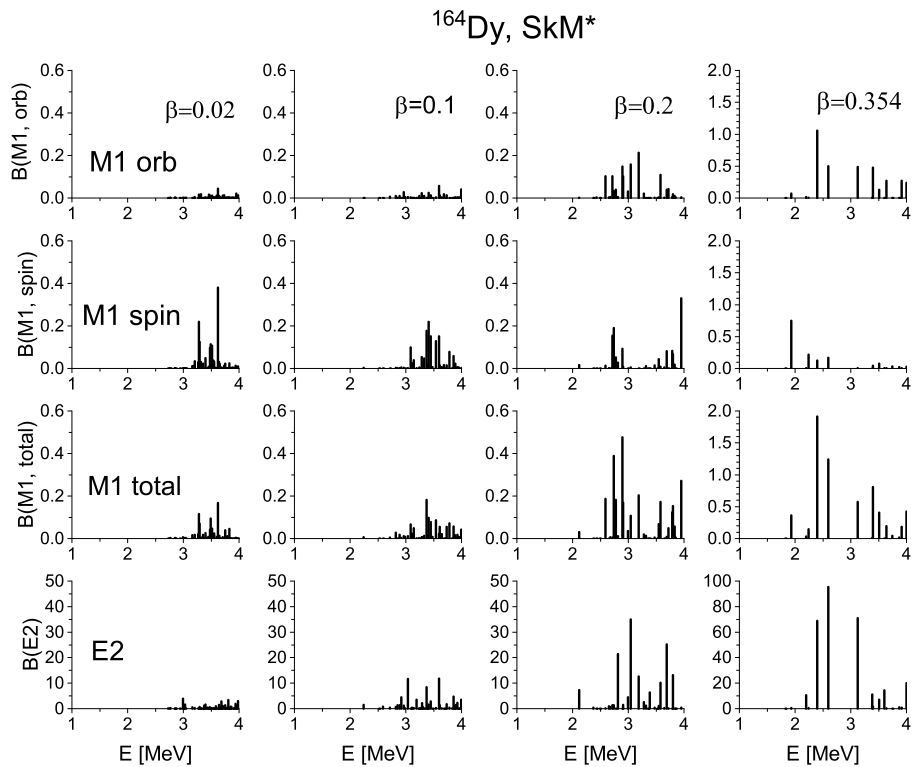}
\label{fig:5}
\caption{Reduced transition probabilities $B(M1,\text{orb})$,  $B(M1,\text{spin})$,
$B(M1,\text{total})$
(in $\mu_N^2$) and $B(E21)$ ( in e$^2$ fm$^4$) in $^{164}$Dy, calculated with the force
SkM* for different quadrupole deformations $\beta$=0.02, 0.1, 0.2, 0.354.}
\end{figure*}

It is instructive to consider distributions of the nuclear current in $1^+$  states.
In Fig. 3, we show average distributions of the proton, neutron, isoscalar
($\Delta T$=0) and isovector ($\Delta T$=1) convective CTD in $^{164}$Dy. Here we average
CTD for all QRPA states at 2.25 - 4 MeV, i.e. at the energy range where most of the
orbital $M1$ strength is concentrated. The averaging allows to smooth individual peculiarities
of the currents of particular QRPA states and thus highlight the main common features of the nuclear
current in the chosen energy range. The procedure of averaging is described in Ref. \cite{Rep13}.

In Fig. 3, the proton current reminds a motion of a fluid contained in the rotating 
ellipsoidal vessel \cite{BM2,Mikh96}.
In this picture, the local flows at the poles and left/right sides have opposite directions.
In our case, the side flows are clockwise and pole flows are anticlockwise. Instead,
in the neutron current, both pole and side flows are anticlockwise. The general flow
is mainly isoscalar (IS) on the poles and isovector (IV) on the sides. In principle,
the IV side flows can be associated with OSR, though this somewhat
contradicts OSR scheme where the most strong flow is expected in the pole regions.

Figure 4 shows spin-up and spin-down currents for the lowest SkM* 2qp spin-flip
state at 1.93 MeV. Both currents have a complicated structure and do not match
smooth SSR-1 currents exhibited in Refs. \cite{Bal_PRC,Bal_PAN}.  Perhaps,
the difference is partly caused by the fact that, unlike the collective SSR-1 states,
QRPA suggests almost 2qp spin-flip states whose currents are not smooth.

\subsection{$^{164}$Dy: spherical limit}

The most disputed point in WFM results is the spin-scissors
treatment of low-energy $1^+$ states, which can be actually realized  only
in deformed nuclei. At the same time, spin-flip states can exit in both deformed and
spherical nuclei. In this connection it is worth to inspect the spherical limit for
low-energy  $1^+$ states. Such WFM analysis was recently performed for $^{164}$Dy
\cite{Bal_PAN}. It was shown that for all three scissors modes (OSR, SSR-1 and SSR-2)
the $M1$ strength fully disappears at the deformation $\delta$=0.135-0.165
($\delta = 0.95\beta$). This result looks doubtful since, if we accept spin-flip
character of SSR-1 and SSR-2, then their $M1$ strength should not fully
vanish in the spherical limit.

This point is inspected in Fig. 5, where  we show SkM* QRPA $M1$ strengths in $^{164}$Dy,
calculated at $\beta$=0.02, 0.1, 0.2 and 0.354 (the last
value is SkM* equilibrium deformation).  We see that, as expected, the low-energy orbital strength  $B(M1,\text{orbit})$ significantly  decreases at $\beta \to$ 0. The quadrupole strength
$B(E21)$ exhibited at the lower panels follow the same trend.
Instead, spin strength $B(M1,\text{spin})$
is downshifted by energy and {\it increases} its integral value. The later is in drastic
contradiction with WFM result \cite{Bal_PAN}. In accordance with Table~\ref{tab-2},
Fig. 5 shows a strong interference of the orbital and spin contributions to $M1$ strength.

\begin{figure}
\includegraphics[width=8cm]{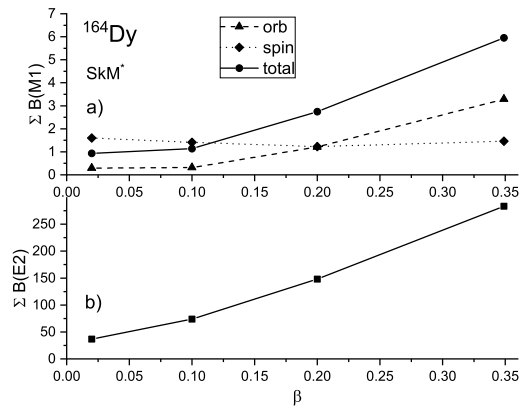}
\label{fig:6}
\caption{a) Orbital (red triangles connected by dashed line), spin (blue rhombuses
connected by dotted  line) and total (black circles  connected by solid line)
summed $B(M1)$ values calculated with the force SkM* for $\beta$=0.02, 0.1, 0.2
and 0.354.  b) The summed $B(E21)$ values for the same deformations.}
\end{figure}

The above trends for$M1$ and $E2$ strengths become even more apparent
in Fig. 6  where these strengths are summed at 0-4 MeV.
We see that the total$M1$ strength is dominated by the spin contribution at
$\beta <0.2$ and by orbital contribution at a higher deformation. The orbital
$M1$ and quadrupole $E2$ strengths vanish at $\beta \to$ 0 while the spin $M1$ strength
even somewhat increases.

\subsection{$^{58}$Ni}

As mentioned above, low-energy spin-flip $M1$ states should exist in both deformed and
spherical nuclei. The latter is demonstrated in Fig. 7 where orbital, spin and total $B(M1)$
values are exhibited for the states at 0-6 MeV in spherical $^{58}$Ni. Actually, in this energy range,
only the lowest $I^{\pi}=1^+$ state at 3.41 MeV is visible (the next state at 4.9 MeV exhibits
very small $M1$ strength and so is almost invisible in the figure). The lowest state is almost
fully (99.5$\%$) exhausted by 2qp spin-flip neutron configuration $ nn[2p_{3/2}, 2p_{1/2}]$
and so is the pure spin-flip excitation. This example shows that,
in contradiction with the WFM conclusion  \cite{Bal_PAN}, low-energy spin-flip states {\it can} exist in
spherical nuclei. This can take place under some obvious conditions: the spin-flip transition
should be of $1ph$-character and connect spin-orbit partners with a low orbital moment. Both these
conditions are fulfilled in  $^{58}$Ni.

Note that other Skyrme forces give similar energies for the lowest
spin-flip $I^{\pi}=1^+$ state in $^{58}$Ni, e.g. 3.42 MeV (SG2) and 3.02 MeV (SVbas).
These values are in a good agreement with the experimental energy  2.902 MeV
of the lowest  $1^+$ state in $^{58}$Ni \cite{database}.

\begin{figure}
\includegraphics[width=8cm]{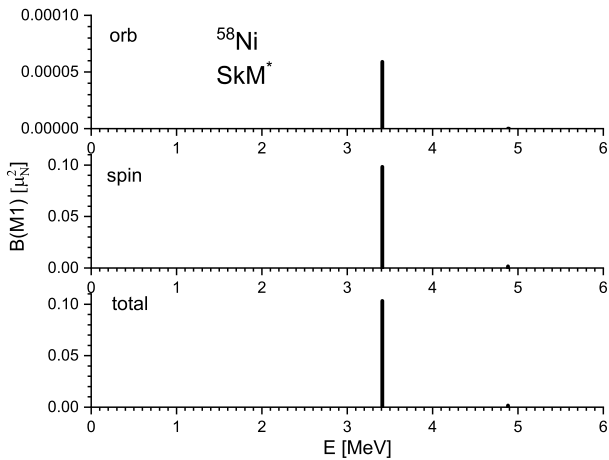}
\label{fig:7}
\caption{Reduced transition probabilities $B(M1)$ in $^{58}$Ni for the orbital, spin and total cases,
calculated with the force SkM*. Note the different scale for the orbital case.}
\end{figure}

\section*{Acknowledgements}
	
The authors thank Prof P.-G. Reinhard for the code SKYAX. A.R. acknowledges support by the Slovak Research and Development Agency under Contract No. APVV-20-0532 and by the Slovak grant agency VEGA (Contract No. 2/0067/21). J.K. appreciates the support by a grant of the Czech Science Agency, Project No. 19-14048S.

\section{Conclusions}

The prediction of the Wigner Function Moment (WFM) method on existence of
low-energy spin-scissors resonances (SSR) in deformed nuclei
\cite{Bal_NPA11,Bal_PRC15,Bal_PRC18,Bal_PRC,Bal_PAN}
was analyzed in the framework of the self-consistent Quasiparticle Random-Phase
Approximation (QRPA) approach. The representative set of  Skyrme forces (SkM*,
SG2, SVbas, and SV-tls) was applied. The main analysis was done for deformed
$^{164}$Dy which, following WFM  predictions, is one of the best candidates
for the search of SSR.

Our calculations show that the lowest $K^{\pi}=1^+$ states in $^{164}$Dy
have a spin-flip character. These states lie at 1.5-2.4 MeV and represent almost
pure two-quasiparticle (2qp) excitations. In principle, these states could be roughly associated
with the predicted SSR. However, there are some arguments against such association.
First, unlike WFM prediction, these states are not collective. Second, their current
distributions significantly differ from WFM SSR currents. Perhaps, the latter is caused
by 2qp character of the lowest spin-flip states, which makes the current distribution complicated
and state-dependent.

In our SkM* calculations, a part of the orbital strength is downshifted to
the SSR region ($E \le $ 2.25 MeV) and, vice versa, the OSR region hosts some
spin-flip strength. As a result, even at the energy range $2.25 $ MeV $< E < $ 4 MeV,
where $M1$ strength is mainly orbital, $1^+$ states demonstrate a
significant interference of the major orbital and minor spin-flip components.

The experimental data \cite{Mar_exp_NRF_1995}
show two distinctive low-energy groups of $1^+$ states in $^{164}$Dy. They are
located at 2.5-2.7 MeV and 3.1-3.2 MeV, respectively. These two groups are treated by WFM
as SSR and OSR. However, in our calculations, spin-flip states lie at $E \le $ 2.4 MeV, i.e.
below the first observed group. Moreover,
following our results, both  groups of the observed $1^+$ states are mainly
produced by fragmentation of the orbital strength.  So, perhaps
these experimental data cannot be considered as the SSR evidence.

The WFM {\it scissors-like} treatment of SSR assumes the nuclear deformation.
Following WFM study \cite{Bal_PAN}, $M1$ strength fully vanishes already at the deformations
$\delta$=0.135-0.165. So it is concluded that SSR cannot exist in spherical nuclei.
Instead, our QRPA calculations for $^{164}$Dy show that spin-flip $M1$ strength remains strong
even in  the spherical limit.
Moreover, we show that the lowest $I^{\pi}=1^+$ state at 3.4 MeV  spherical $^{58}$Ni is
is spin-flip one. So, in contradiction with WFM conclusions, low-energy $M1$ states of the spin
character {\it can} exist in spherical nuclei.

Altogether, one may conclude that our present QRPA calculations do not confirm the existence of SSR.
Perhaps, more studies are necessary to draw the final conclusions. It would be useful to include to
the self-consistent microscopic analysis the coupling with complex configurations. As for WFM,
the strange result with disappearance of $M1$ strength in the spherical limit should be checked.
Besides, to our knowledge,  the giant spin-flip $M1$ resonance does not appear in WFM calculations,
which make questionable any analysis of spin modes within this model. Further, WFM calculations
give in $^{164}$Dy the 1.47-MeV $K^{\pi}=1^+$ state with a huge $B(E2)$ value \cite{Bal_PRC}. Such
 state is absent in both experiment and numerous microscopic calculations.

 Finally note that low-energy spin-flip $M1$ states can be sensitive to tensor forces.
 In our calculations for $^{164}$Dy, the effect of tensor force is modest. Anyway,
low-energy spin-flip $M1$ states are in general very promising object for investigation
of the impact of tensor forces.

\end{document}